\documentclass[twocolumn]{aa}
\usepackage{graphicx}
\usepackage{txfonts}
\begin{document}
\titlerunning{Structural Parameters of NGC 5128 Globular Clusters}
   \title{Structural Parameters from Ground-based Observations of
Newly Discovered Globular Clusters in \object{NGC\,5128}}

   \author{M. G\'omez
          \inst{1}
          \and
          D. Geisler
	  \inst{1}
	  \and
	  W.~E. Harris
	  \inst{2}
	  \and
	  T. Richtler
	  \inst{1}
	  \and
	  G.~L.~H. Harris
	  \inst{3}
	  \and
	  K.~A. Woodley
	  \inst{2}
          }

   \offprints{M. G\'omez}

   \institute{Grupo de Astronom\'{\i}a, Depto. de F\'{\i}sica, Universidad de Concepci\'on,
              Casilla 160-C, Concepci\'on, Chile.
              \email{matias@astro-udec.cl}
         \and
             Department of Physics and Astronomy,
	     McMaster University, Hamilton ON L8S 4M1, Canada.
	 \and
	     Department of Physics, 
	     University of Waterloo,
	     Waterloo, Ontario, N2L 3G1, Canada.
             }

   \date{Received; accepted}

   \abstract{

We have investigated a number of globular cluster candidates from a recent wide-field 
study by Harris et~al.~(\cite{harris04a}) of the giant elliptical galaxy \object{NGC\,5128}.
We used the Magellan I telescope + MagIC camera under excellent seeing conditions 
($0.3\arcsec-0.6\arcsec$) and obtained very high resolution images for a sample of 44 candidates. 
Of these, 15 appear to be bonafide globular clusters in \object{NGC\,5128} while the rest are either 
foreground stars or background galaxies. We also serendipitously discovered 18 new 
cluster candidates in the same fields. Our images allow us to study the light 
profiles of the likely clusters, all of which are well resolved. 
This is the first ground-based study of structural parameters for globular clusters
outside the Local Group. We compare the psf-deconvolved profiles with King models 
and derive structural parameters, ellipticities and surface brightnesses. 
We compare the derived structural properties with those of other well-studied
globular cluster systems. In general, our clusters are similar in size, ellipticity, 
core radius and central surface brightness to their counterparts in other galaxies, 
in particular those in \object{NGC\,5128} observed with HST by Harris et~al.~(\cite{harris02}). 
However, our clusters extend to higher ellipticities and larger half-light
radii than their Galactic counterparts, as do the Harris et~al. sample. Combining 
our results with those of Harris et~al. fills in the gaps previously existing in $r_h - M_V$
parameter space and indicates that any substantial difference between presumed distinct 
cluster types in this diagram, including for example the Faint Fuzzies of Larsen 
\& Brodie~(\cite{larsen00}) and the `extended, luminous' M31 clusters of Huxor 
et~al.~(\cite{huxor05}) is now removed and that clusters form a continuum in this diagram. Indeed, 
this continuum now extends to the realm of the Ultra Compact Dwarfs.
The metal-rich clusters in our sample have half-light radii that are almost twice as 
large in the mean as their metal-poor counterparts, at odds with the generally accepted 
trend. The possibility exists that this result could be due in part to contamination by
background galaxies. We have carried out additional analysis to quantify this contamination.
This shows that, although galaxies cannot be easily told apart
from clusters in some of the structural diagrams, the combination of excellent image quality 
and Washington photometry should limit the contamination to roughly 10\% of the population of
cluster candidates. 
Finally, our discovery of a substantial number of new cluster candidates in the 
relatively distant regions of the \object{NGC\,5128} halo suggests that current values of the 
total number of globular clusters may be underestimates.

   }

   \maketitle

\section{Introduction}

At a distance of $\sim 4$ Mpc (Soria et~al.~\cite{soria96}, Rejkuba~\cite{rejkuba04}, 
Harris et~al.~\cite{harris04a}), \object{NGC\,5128} (Centaurus A)
is the closest giant elliptical (gE) galaxy (see Israel~\cite{israel98} for a review). 
It possesses a rather low specific frequency of globular clusters (GCs), with
$S_N=1.4\pm0.2$ (Harris et~al.~\cite{harris04b}, hereafter HHG), yet still hosts around
1000 GCs. This makes its globular cluster system (GCS) the largest of any
galaxy known within $\sim 15$Mpc. It is therefore a prime  target for studies of
extragalactic GCSs. This is especially important given that systematic
differences are suspected among GCSs depending on their host galaxy type and environment 
(Fleming et~al.~\cite{fleming95}, Kissler-Patig~\cite{kissler00}). 
\object{NGC\,5128} offers a unique opportunity to study in great detail the GCS of a gE and to 
both compare it with closer systems, hosted only by late-type and dwarf galaxies, and 
to use it as a prototype for the GCSs of more distant gEs.

However, the study of the \object{NGC\,5128} GCS has been hampered by a set of 
observational circumstances which make further study far from straightforward. The
low galactic latitude ($b=+19^{o}$) makes the contamination by foreground stars a major issue. These,
together with background galaxies, vastly outnumber the cluster population and 
many of them occupy a similar range in colour and magnitude, even if using a colour 
like the Washington $C-T_{1}$ index, which has proven especially powerful in distinguishing 
clusters from contaminating objects (Geisler et~al.~\cite{geisler96b}, Dirsch et~al.~\cite{dirsch03}). 
In their wide-field Washington photometric investigation of \object{NGC\,5128}, Harris et~al.~(\cite{harris04a}) 
estimated that bonafide GCs constitute only $\sim1$\% of the $10^5$ objects they observed.
In addition, \object{NGC\,5128} is so close that its GCS is very spread out in angular size, and some 
clusters have been found at distances as large as $40\arcmin$ from the optical center 
(Harris et~al.~\cite{harris92}, Peng et~al.~\cite{peng04}), requiring the use of very wide 
field of view detectors for a comprehensive study. Yet it is distant enough that GCs 
cannot be easily told apart from the background and foreground population via 
image resolution, at least with typical ground-based images. With characteristic 
half-light radii of $0.3\arcsec - 1.0\arcsec$ (Harris et~al.~\cite{harris02}), 
excellent seeing conditions are needed to resolve the majority of the clusters 
from the ground.

The study of the structure of globular clusters has led to the discovery that they define a ``fundamental plane''
in analogy to that of elliptical galaxies. This is, they occupy a narrow region in multi-parameter space. This
has been shown for Milky Way clusters (Djorgovski~\cite{djorgovski95}) as well as for a few other GCSs in the Local
Group (Djorgovski et~al.~\cite{djorgovski97}, Barmby et~al.~\cite{barmby02}, Larsen et~al.~\cite{larsen02})
and a sample of \object{NGC\,5128} GCs studied with HST by Harris et~al.~(\cite{harris02}).
As the structure of a cluster is the
result of its dynamical history, it is of great importance to compare cluster structures from a variety of galaxies
along the Hubble sequence and to look for correlations with galaxy type. This is especially true in the case of 
elliptical galaxies, which have presumably a more complex formation history and are likely to have experienced 
several distinct formation events, as suggested by the usual bimodality in the colours of their GCSs 
(Geisler et~al.~\cite{geisler96b}, Kundu \& Whitmore~\cite{kundu01}, 
Larsen et~al.~\cite{larsen01b}). The half-light radius $r_{h}$ should remain roughly
constant throughout the life of a GC (e.g. Spitzer \& Thuan~\cite{spitzer72}, Aarseth 
\& Heggie~\cite{aarseth98}), so its current size should reflect conditions of the
proto-GC cloud. Any systematic variation of $r_{h}$ within or among galaxies
can provide insights into GC formation. For example, Mackey \& Gilmore~(\cite{mackey04}) found
very different $r_{h}$ distributions for disk/bulge, old halo and young halo Galactic GCs.
Therefore, studying structural parameters of more 
GCSs and especially those of gE galaxies may help our understanding of galaxy formation.

In addition, a number of cluster subtypes have been suggested recently
on the basis of their distinct properties. For example, Larsen \& Brodie~(\cite{larsen00}) 
find a class of extended, intermediate luminosity clusters in NGC\,1023 which 
they refer to as 'Faint Fuzzies'. They also find similar objects in NGC\,3384 
(Larsen et~al.~\cite{larsen01b}). Since these objects have so far only been identified
in these two lenticular galaxies, it has recently been suggested (Burkert 
et~al.~\cite{burkert05}) that these objects form {\em only\/} in such galaxies and indeed that their
formation may be intimately related to that of their parent galaxy. Similarly,
Huxor et~al.~(\cite{huxor05}) have discovered 3 very large, luminous GCs in the halo
of M31 which appear to be unique. In addition, a new type of Ultra Compact
Dwarf (UCD) galaxy now appears to exist (Hilker et~al.~\cite{hilker99}, Drinkwater 
et~al.~\cite{drinkwater02}) which may or may not be related to GCs (Mieske et~al.~\cite{mieske02}). 
Ha\c{s}egan et~al.~(\cite{hasegan05}) report the discovery of several bright
objects in the Virgo Cluster which they refer to as DGTOs (Dwarf-Globular Transition Objects)
and propose the use of the M/L ratio to distinguish between bright GCs and UCDs.
How unique are such objects? Do they exist in other galaxies, of other types? 
Are their properties truly distinct from those of other GCs or do GCs populate 
a continuum with no clear subclasses? What is the relation of GCs to the UCDs, 
if any? Such questions can be addressed by obtaining high quality structural 
parameters for as many different GCs in as many different types of galaxies as possible.

There have been intriguing hints (Kundu \& Whitmore~\cite{kundu01}, Larsen et~al.~\cite{larsen01b}, 
Larsen \& Brodie~\cite{larsen03}) that the blue clusters in gEs are systematically 
larger by some $20\%$ on average than their red counterparts, based on WFPC2 data. 
Recent Virgo Cluster ACS data (Jord\'an et~al.~\cite{jordan05}) strengthen this result.
It is still not clear how wide-spread this effect is and what its cause may be. 
For example, it has been suggested that the effect may stem from real differences in 
the formation and evolution of these distinct cluster subpopulations (Jordan~\cite{jordan04}) 
or may simply reflect the fact that the red clusters are generally more centrally 
concentrated than their blue companions (e.g. Geisler et~al.~\cite{geisler96b}) and that 
the larger tidal forces there lead to more compact clusters in the inner regions 
(Larsen \& Brodie~\cite{larsen03}).

However, to date little is known about the structural parameters 
of GCs in gEs. 
The only exception is the \object{NGC\,5128} GCS.  
Using WFPC2 data, Harris et~al.~(\cite{harris02}) obtained
images for 27 GCs and derived their structural parameters. Combining with similar data
for inner halo clusters from Holland et~al.~(\cite{holland99}), they found that
the light profiles fit classic King models very well and that their structural
parameters were similar to those of MW GCs, although their ellipticities were
substantially larger than those of MW GCs and much more like those of M31 
clusters.

Recently, Martini \& Ho~(\cite{martini04}) have obtained the velocity dispersions for the brightest 14 
clusters in \object{NGC\,5128}.
Combining these data with the Harris et~al.~(\cite{harris02}) structural parameters, 
they were able to construct the fundamental plane for the clusters and showed that they follow approximately 
the same relationships found for Local Group clusters. This, in spite of their
extreme masses and luminosities (about 10 times larger than nearby counterparts).
However, since the discovery of the first GC in \object{NGC\,5128} (Graham \& Phillips~\cite{graham80}), its known GC       
population has steadily increased and is estimated today to be $\sim1000$ (Harris et~al.~\cite{harris04b}), 
so one of course would like to extend such analysis towards less luminous clusters and study a much more 
representative sample before definitive conclusions can be reached.

In this paper, we report the observation of a number of small fields around \object{NGC\,5128} under exceptional 
seeing obtained at the Magellan I telescope. These images allowed us to resolve known cluster candidates 
(and thus confirm or discard their cluster nature on the basis of their resolution and shape)
and to detect a number of new ones. We have used these high-resolution images to derive structural parameters,
surface brightnesses and central mass densities, which we compare to those of other well-studied GCSs. 
These are the first structural parameters derived for GCs beyond the Local Group using ground-based images.

The paper is organised as follows. In Section 2, we present the observations and reductions 
and describe our procedure for identifying clusters. In Section 3, we derive the structural parameters 
by fitting to models. In Section 4, we discuss our results and compare the derived structural parameters with
those of other GCSs. We discuss the contamination by background galaxies and its effects in Section 5.
Finally, we summarize our major findings in Section 6.

\section{The data}

\subsection{Observations}

The fields were selected from the sample of Harris et~al.~(\cite{harris04a}). 
They present a list of 327 candidate GCs based on colour and incipient resolution on 
their CTIO 4m BTC (Big Throughput Camera) frames with seeing $\sim1.2\arcsec$ and pixel 
scale of $0.42\arcsec$/pix. On the basis of brightness, we selected 30 fields, 
which contain about 45 of these candidates
within the field of view of MagIC (Magellan Instant Camera). 
We concentrated on clusters in the distant halo of the galaxy in order to
explore their nature in greater detail. These candidates had  not been observed in 
previous spectroscopic work, thus their true nature was unknown.

\begin{figure}
   \centering
    \includegraphics[width=9cm]{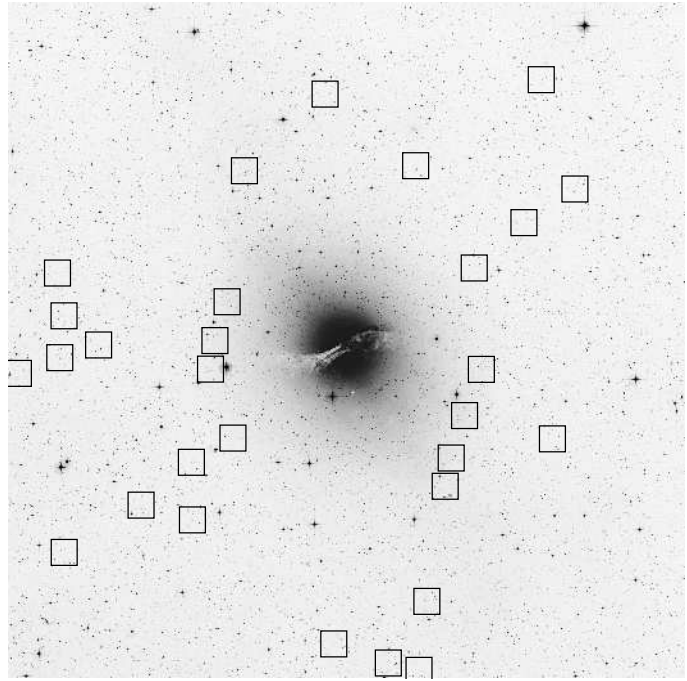}
      \caption{A one square-degree image from DSS. Overlaid are 29 of the 30 fields observed
with MagIC (one is slightly off this FOV) in actual size (2.3'$\times$2.3').
The orientation is N up, E to the left.}
         \label{fig.n5128_dss}
   \end{figure}

Images were obtained with the Magellan~I 6.5m telescope at Las Campanas Observatory
on Jan. 30, 31 and Feb. 1, 2004 with MagIC.
This is a 2K $\times$ 2K CCD with a scale of $0.069\arcsec$/pix (very 
similar to that of the PC on WFPC2), spanning a field of view of 
$2.3\arcmin \times 2.3\arcmin$. All of the fields (overlaid in Fig.~\ref{fig.n5128_dss}) 
were observed at least once in $R$ and a few of them also in $B$. For these latter 
observations (which were taken during the third night) the seeing was significantly 
worse (over $0.7\arcsec$). Therefore, our results refer only to the $R$ frames, in which
we enjoyed superb seeing of $0.3 - 0.6\arcsec$.
Typical exposure times were  180-300 sec.
Tables~1 and 2 give details of the observations and observed clusters.
The nights were not photometric but none of our results depends on absolute
photometry acquired during this run.

\subsection{Data reduction}

The frames were bias-subtracted and flatfield corrected using the 
IRAF\footnote{IRAF is distributed by the National Optical Astronomy Observatories,
    which are operated by the Association of Universities for Research
    in Astronomy, Inc., under cooperative agreement with the National
    Science Foundation.} script
MagIC-tools, as described by Phillips~(\cite{phillips02}). For a few cases, the fields were observed twice. 
Those frames with comparable seeing were combined using {\it imcombine\/} after the alignment 
with {\it geomap\/} and {\it geotrans\/} with respect to the frame with higher S/N. No cosmic 
rays were detected near or on the cluster candidates, hence no attempt was made to remove 
them. Bad pixels were replaced by interpolation with neighbouring pixels using the task 
{\it fixpix\/}.

\subsection{Cluster identification}

The globular clusters in \object{NGC\,5128} have characteristic half-light radii
of $0.3''-1''$ (measured from HST imaging, Harris et~al.~\cite{harris02}), and so with sub-arcsecond 
seeing quality the great majority of the clusters should be distinguished from stars
and large and/or irregular and/or elliptical galaxies.
Given our excellent seeing and
the high resolution of MagIC, we were not only able to achieve this but also
discovered many new faint candidates that serendipitously lie in the same fields.
This was done by subtracting the stellar PSF from
all sources in each frame and visually inspecting the residuals. DAOPHOT II (Stetson~\cite{stetson87})
was run under IRAF first to detect the sources with the {\it daofind\/} algorithm. Aperture
photometry was then performed with a radius of 4 pixels. This was the basis to build the PSF
by using $\sim$30 bright, isolated stars per frame. Cluster candidates had broader profiles
and we are confident that they were not used for the PSF. Anyway, the large number of
stars used for the PSF would blur any effect of including possible compact clusters into 
this category.

\begin{figure}
   \centering
    \includegraphics[width=8.8cm]{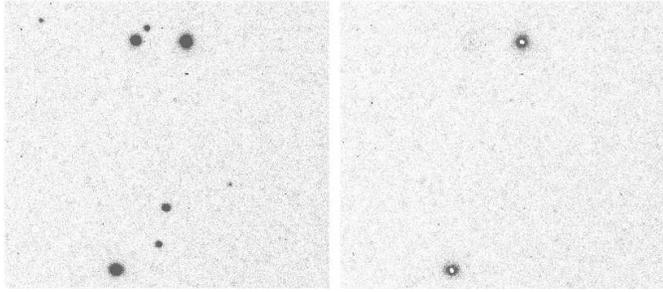}
      \caption{llustration of how the resolution technique works in discriminating
resolved globular clusters in \object{NGC\,5128} from unresolved stars and also
background galaxies. On the left is a small area ($37\arcsec \times37\arcsec$) of one of our MagIC images. 
On the right, one sees that after subtraction of a stellar psf, stars disappear while globular
clusters (two of which are shown) appear as round doughnuts with oversubtracted 
centers and undersubtracted wings.}
         \label{fig.psf_subtraction}
   \end{figure}

The non stellar appearance of the cluster candidates allowed a straightforward identification.
After subtraction of the PSF, resolved objects (both \object{NGC\,5128} GCs as well as
certain types of
background galaxies) leave a ``doughnut-shape'', being undersubtracted in the 
wings and oversubtracted in the center, as shown in Fig.~\ref{fig.psf_subtraction}.
Resolved objects were further culled by eliminating those that had very
large and/or irregular  profiles in order to weed out as
many galaxies as possible. However, relatively small, regular, round galaxies
will not be recognized and may contaminate our sample.

After a visual inspection of the resulting images, we found that, starting from 44 cluster
candidates, only 17 of them (39\%) had cluster-like residuals. Twelve of the 44 were not resolved and show a
star-like profile, and 15 are background galaxies. This illustrates the general severe problem
of field contamination discussed above:  even after careful selection of candidates
from their colors and (in some cases) barely extended appearance on good-quality
ground-based images, the total number of non-cluster objects in the list is larger than 
the cluster population we are trying to find.

In addition, we could identify several new fainter candidates in our MagIC fields, typically 2-3
per frame. For the majority of them, Washington photometry already exists from 
Harris et~al.~(\cite{harris04a}) and was used to apply the same color selection 
as they used to generate their candidate list. In all, 48 new candidates were found.
Of these, 15 have very red colours and are presumably background galaxies 
and 11 had no photometry available. 
This happened either because they were too faint or they were located in the gaps between the BTC 
frames, where the photometry data come from. Thus, 22 new, good cluster 
candidates have been added to the analysis.
The comparison with previous work shows that three of them were already in the confirmed GC 
catalog of Peng et~al.~(\cite{peng04}), and one is listed as a GC by Hesser et~al.~(\cite{hesser84}). 
They are marked in Tables 2 and 3 as `PFF' and `C' objects respectively and hence the number
of truly new clusters is 18.

This work therefore gives a total of 39 high-quality candidate clusters, 18 new 
and 21 previously existing candidates.
After our imaging observations were procured, Woodley et~al.~(\cite{woodley05}) obtained spectra
for an independent sample of nearly a hundred of the HHG GC candidates and were able to classify them as 
bonafide \object{NGC\,5128} GCs, foreground stars or background galaxies on the basis
of their radial velocities. Nine objects are in common between the two studies and are
accordingly labelled in Table 1.
Four of the objects identified as GCs by us are 
also GCs from their radial velocities, while two objects we classified as GCs
(\#127 and \#94) are actually distant galaxies according to radial velocity (and 
labelled 'WHH gal.' in Table 1). In addition, there is 
perfect agreement in the (independent) classification for three foreground stars (see Table~1.)
We were not able to perform this comparison for
our newly discovered cluster candidates since no other data exists for them
besides the Harris et~al.~(2004a) photometry other than the fact that four of
them were independently classified as GCs based on lower resolution images.
This test indicates that galaxies still contaminate our final GC sample, as 
expected. A much more detailed estimate of galaxy contamination is given in Section~5.

Although we have eliminated the two galaxies 94 and 127 from further consideration,
both of them provide interesting properties.
Object 127 does not deviate significantly from a typical King profile and all derived parameters
(as well as photometry) are in the range of GCs. Perhaps its small core and effective radii could be the
only indication of its strangeness. Object 94 was taken for a cluster in projection next 
to a distant galaxy but
from its velocity, it might consist of a pair of galaxies, separated by $\sim1.6''$.
In Table 2 we present observational data for our new cluster candidates, labeled 
with an 'N' prefix, and present the full list of our final cluster candidates.

\begin{table*}
\caption{Washington photometry and classification of the 44 observed targets
(from Harris et al (2004a). The first column gives their
identification. Objects in common 
with Woodley et~al.~(\cite{woodley05})
are indicated by WHH and their
classification (star/galaxy) or cluster number. The coordinates ($\alpha$,$\delta$)
are given in the second and third column. The colours in the Washington system 
from Harris et al. (2004a) are listed in the following three columns.
The last column indicates our classification, according to their morphology in our MagIC frames.}
\label{tab.list}
\centering
\begin{tabular}{l c c c c c l }
\hline\hline
object ID & $\alpha_{2000}$ & $\delta_{2000}$ & $T_{1}$ & $M-T_{1}$ & $C-T_{1}$ & Classification \\
(HHG)	  & 		    & 		         &         &           & 	        &                          \\
\hline

017 (WHH star)	& 201.586426 & -43.15366 & 17.447	&  0.644 & 1.176  &	star			\\

021		& 201.627686 & -43.01647 & 17.546	&  0.568 & 1.433  &	star			\\
022 (WHH001)	& 201.089172 & -43.04356 & 17.630	&  0.704 & 1.484 & 	cluster + star		\\

032		& 200.909683 & -42.77305 & 17.984	&  0.740 & 1.462 & 	cluster			\\
034		& 201.629135 & -43.01801 & 18.001	&  0.636 & 1.524  &	star			\\

036 (WHH star)	& 201.010620 & -42.82483 & 18.046	&  0.724 & 1.586 & 	star			\\
037		& 200.983215 & -42.61264 & 18.083	&  1.134 & 2.588 & 	star			\\
038 (WHH star)	& 201.637817 & -43.05389 & 18.092	 & 0.526 & 1.088  &	star			\\
039		& 201.670975 & -43.27308 & 18.108	&  0.803 & 1.563  &	star			\\

049		& 201.773285 & -43.25498 & 18.365	&  0.923 & 2.461  &	galaxy			\\
050		& 200.926392 & -43.16050 & 18.490	&  0.791 & 1.630 & 	star			\\
051 (WHH004)	& 201.169159 & -43.22168 & 18.531	&  0.868 & 1.927 & 	cluster 		\\

060		& 201.638031 & -43.04905 & 18.667	&  1.078 & 2.082  &	star			\\
068		& 201.231567 & -42.74451 & 18.765	&  0.743 & 1.770 & 	galaxy			\\
069		& 201.639236 & -43.01849 & 18.770	&  0.541 & 1.198  &	star			\\
074		& 202.025970 & -43.06270 & 18.894	&  1.074 & 2.653  &	galaxy			\\
080		& 201.435043 & -42.64040 & 18.972	&  1.048 & 2.754 & 	cluster			\\
081		& 201.134018 & -43.18247 & 18.992	&  0.764 & 1.470 & 	galaxy			\\

084		& 200.956879 & -43.12678 & 19.023	&  0.885 & 2.166 & 	galaxy			\\
086 (WHH031)	& 201.672623 & -43.19029 & 19.048	&  0.846 & 1.618  &	cluster 		\\
093		& 201.606171 & -42.95170 & 19.173	&  0.801 & 1.792  &	cluster			\\
094 (WHH gal.)	& 201.117294 & -42.88461 & 19.191	&  0.978 & 1.965 & 	cluster + galaxy	\\
098		& 201.188309 & -43.38884 & 19.255	&  0.824 & 1.660 & 	galaxy			\\

099		& 201.099915 & -42.90296 & 19.266	&  0.679 & 1.270 & 	galaxy 			\\
102		& 201.263916 & -43.48138 & 19.351	&  0.916 & 2.200 & 	cluster 		\\
104		& 201.942627 & -43.03859 & 19.374	&  0.921 & 1.943  &	cluster 		\\
105		& 201.576157 & -42.75731 & 19.377	&  0.879 & 1.865  &	galaxy			\\
106 (WHH029)	& 201.591995 & -43.15297 & 19.379	&  0.788 & 1.538  &	cluster			\\
120		& 201.935272 & -42.97654 & 19.521	&  0.806 & 1.799  &	galaxy			\\
127 (WHH gal.)	& 201.144211 & -43.21404 & 19.588	 & 0.949 & 1.932 & 	cluster 		\\
128		& 201.100601 & -42.90571 & 19.598	&  0.903 & 1.939 & 	cluster			\\
129		& 201.949738 & -42.91334 & 19.600	&  0.972 & 2.599  &	cluster 		\\
130		& 202.090408 & -42.90478 & 19.602	&  0.753 & 1.649  &	galaxy			\\
141		& 201.157684 & -43.16919 & 19.720	&  1.000 & 2.053 & 	cluster			\\
145		& 201.120895 & -43.11254 & 19.776	&  0.835 & 1.377 & 	galaxy 			\\
147		& 201.864517 & -43.01892 & 19.794	&  0.670 & 1.197  &	galaxy			\\
200		& 201.201157 & -43.47463 & 20.398	&  0.656 & 1.460 & 	star			\\
208		& 201.398788 & -42.64296 & 20.467	&  1.176 & 2.235  &	star			\\
210		& 201.201355 & -43.47885 & 20.485	&  0.974 & 2.225 & 	cluster 		\\
225		& 201.375885 & -43.45494 & 20.725	 & 1.001 & 1.745 & 	cluster			\\
228		& 201.665817 & -43.27797 & 20.740	&  1.009 & 1.819  &	galaxy			\\
244		& 201.200668 & -43.50775 & 21.004	&  0.799 & 1.390 & 	galaxy			\\
246		& 201.927948 & -43.32676 & 21.017	&  1.028 &        &	galaxy			\\
327		& 201.620071 & -43.00220 &              &        & 	&	cluster + star		\\

\hline
\end{tabular}
\end{table*}

\begin{table*}
\caption{Washington photometry and observational details for all of our final GC
candidates. Clusters with no prefix are candidates from Harris et~al.~(\cite{harris04a}) database (from Table 1). Those
labelled as 'N' are newly identified objects in this study (see text for detail) WHH objects are those also present
in Woodley et~al.~(\cite{woodley05}). 
Note that clusters 022 and N3
were observed twice.}
\label{tab.observations}
\centering
\begin{tabular}{l c c c c c c l c }     
\hline\hline       
cluster ID & $\alpha_{2000}$ & $\delta_{2000}$ &  	T$_{1}$    &	$C-T_{1}$ & date & Airmass & Exp. time & Seeing  \\
(HHG)	   & 		 	 & 		 &              &              & 	 &         &  (sec.)    & ('')    \\
\hline                    

022 (WHH001)  &	201.089172 &	-43.043560 &	17.630  &	1.484 &	30.Jan & 1.08	 &	180	& 0.32	\\
022 (WHH001)  &	201.089172 &	-43.043560 &	17.630  &	1.484 &	30.Jan & 1.07	 &	300	& 0.35	\\
032  	& 	200.909683 &	-42.773048 &	17.984  &	1.462 &	30.Jan & 1.14	 &	300	& 0.38	\\
051 (WHH004)  &	201.169159 &	-43.221680 &	18.531  &	1.927 &	30.Jan & 1.05	 &	180	& 0.38	\\
080   	& 	201.435043 &	-42.640400 &	18.972  &	2.754 &	31.Jan & 1.14 	 &	180	& 0.53	\\
086 (WHH031)  &	201.672623 &	-43.190289 &	19.048  &	1.618 &	31.Jan & 1.09	 &	180	& 0.59	\\
093   	& 	201.606171 &	-42.951698 &	19.173  &	1.792 &	31.Jan & 1.11	 &	180	& 0.54	\\
102  	& 	201.263916 &	-43.481380 &	19.351  &	2.200 &	30.Jan & 1.04	 &	300	& 0.35	\\
104  	& 	201.942627 &	-43.038589 &	19.374  &	1.943 &	31.Jan & 1.06	 &	180	& 0.44	\\
106 (WHH029) &	201.591995 &	-43.152969 &	19.379  &	1.538 &	31.Jan & 1.12	 &	180	& 0.50	\\
128  	& 	201.100601 &	-42.905708 &	19.598  &	1.939 &	30.Jan & 1.07	 &	180	& 0.30	\\
129  	& 	201.949738 &	-42.913342 &	19.600  &	2.599 &	31.Jan & 1.06	 &	240	& 0.43	\\ 
141  	& 	201.157684 &	-43.169189 &	19.720  &	2.053 &	30.Jan & 1.05	 &	180	& 0.41	\\
210  	& 	201.201355 &	-43.478851 &	20.485  &	2.225 &	30.Jan & 1.04	 &	180	& 0.39	\\
225  	& 	201.375885 &	-43.454941 &	20.725  &	1.745 &	31.Jan & 1.15	 &	180	& 0.57	\\
327  	& 	201.620071 &	-43.002201 &	        &	      & 31.Jan & 1.11	 &	180	& 0.53	\\
N1 (C40)    &	200.926392 & 	-43.160500 &	18.490  &	1.630 & 30.Jan & 1.12	 &	180	& 0.34 \\
N3     &	201.072800 & 	-43.039169 &	20.897  &	1.848 & 30.Jan & 1.08	 &	180	& 0.32 \\
N3     &	201.072800 & 	-43.039169 &	20.897  &	1.848 & 30.Jan & 1.07	 &	300	& 0.35 \\
N5     &	201.136230 & 	-43.094910 &	20.774  &	1.680 & 30.Jan & 1.06	 &	300	& 0.30 \\
N7 (PFF06)  &	201.098740 & 	-43.131149 &	18.731  &	1.373 & 30.Jan & 1.06	 &	300	& 0.30 \\
N8     &	201.103271 & 	-43.119518 &	20.057  &	1.654 & 30.Jan & 1.06	 &	300	& 0.30 \\
N9     &	201.116486 & 	-43.109959 &	21.933  &	1.871 & 30.Jan & 1.06	 &	300	& 0.30 \\
N10 (PFF09) &	201.130554 & 	-43.190781 &	19.258  &	1.474 & 30.Jan & 1.05	 &	180	& 0.41 \\
N11    &	201.172958 & 	-43.214771 &	20.296  &	1.464 & 30.Jan & 1.05 	 &	180	& 0.38 \\
N12    &	201.140121 & 	-43.200439 &	20.119  &	1.568 & 30.Jan & 1.05 	 &	180	& 0.38 \\
N21    &	201.437943 & 	-42.649410 &	19.845  &	2.331 & 31.Jan & 1.14 	 &	180	& 0.53 \\
N23    &	201.561600 & 	-42.757729 &	20.072  &	0.815 & 31.Jan & 1.13	 &	240 	& 0.49 \\
N24    &	201.598083 & 	-43.151272 &	21.290  &	1.691 & 31.Jan & 1.12	 &	180	& 0.50 \\
N25    &	201.571945 & 	-43.166100 &	20.914  &	2.207 & 31.Jan & 1.12	 &	180	& 0.50 \\
N26 (PFF092) & 201.588715 & 	-42.955292 &	19.458  &	1.727 & 31.Jan & 1.11	 &	180	& 0.54 \\
N30    &	201.675735 & 	-43.286640 &	19.916  &	1.247 & 31.Jan & 1.09	 &	180	& 0.57 \\
N32    &	201.691513 & 	-43.194920 &	20.034  &	1.366 & 31.Jan & 1.09	 &	180	& 0.59 \\
N33    &	201.686783 & 	-43.204910 &	20.156  &	1.887 & 31.Jan & 1.09	 &	180	& 0.59 \\
N34    &	201.766739 & 	-43.259491 &	20.267  &	1.626 & 31.Jan & 1.08	 &	180	& 0.60 \\
N35    &	201.787048 & 	-43.271809 &	21.005  &	1.957 & 31.Jan & 1.08	 &	180	& 0.60 \\
N37    &	201.854248 & 	-43.002270 &	21.318  &	1.993 & 31.Jan & 1.08	 &	240	& 0.52 \\
N41    &	201.974548 & 	-42.922680 &	19.988  &	2.515 & 31.Jan & 1.06	 &	240	& 0.43 \\
N42    &	201.955078 & 	-42.932850 &	18.436  &	2.591 & 31.Jan & 1.06	 &	240	& 0.43 \\
\hline
\end{tabular}
\end{table*}

\section{Analysis}

\subsection{Fit to the light profiles}

All of our final cluster candidates have FWHMs significantly larger than those of stars, thus the
residuals are noticeable. No identification was attempted for faint objects, as they were beyond the
limiting magnitude of the Harris et~al.~(\cite{harris04a}) database
($R\sim 22$ mag).

For all objects listed in Table~2, a 2-D fitting was performed to derive their morphological parameters after
deconvolution with the stellar PSF. This was done  using the task {\it ishape\/} under 
BAOLAB\footnote{BAOLAB is available at http://www.astro.ku.dk/$\sim$soeren/baolab} (Larsen~\cite{larsen99}).
In two cases, a star was located $\sim 2-3''$ away from the cluster candidate. They were first removed 
using PSF-fitting and then {\it ishape\/} was run on the clean image.

Ishape deconvolves the observed light distribution with the PSF and then performs a 2-D fitting 
by means of a weighted least-square minimization and assuming
an intrinsic profile. For the latter we have chosen a King profile (King~\cite{king62}), which has the form:

\begin{equation}
\mu(r) = k \Bigl[\frac{1}{\sqrt{1 + r^{2}/r_{c}^{2}}} - \frac{1}{\sqrt{1 + r_{t}^{2}/r_{c}^{2}}}\Bigr]^{2} ~~, r < r_{t}
\end{equation}

Ishape also offers other profiles. Among them, the `Moffat-type' is  the most 
frequently used (see Elson et~al.~\cite{elson87}, 
Larsen~\cite{larsen01a}, Mackey \& Gilmore~\cite{mackey03}). Although we extensively tried different 
models and parameters, the best results were obtained with King profiles, except for a very few cases 
which were better fit by a Moffat model and which are presumably background galaxies (see discussion).

King profiles are known to provide excellent fits to Milky Way GCs and are characterised 
by a tidal radius $r_{t}$ and a core radius $r_{c}$. Alternatively, a concentration parameter $c=r_t/r_c$ 
can be defined, which we use here. Note that this definition differs from the more familiar 
$c=\mathrm{log}(r_t/r_c)$. We prefer
to use the first definition to keep consistency with the outputs from {\it ishape\/}. The $c$ parameter 
can be kept fixed or allowed to vary during the fitting process and is the most uncertain of the fitted
parameters (Larsen~\cite{larsen01a}). However, a number of tests with our images show that, in fact,
$c$ is stable typically within 20\% when the radius of the fit increases from 15 to 25 pixels. This
is encouraging given the large errors found in the literature. On the other hand, one should keep in mind the
corresponding uncertainty in the tidal radii, which are derived directly from $c$ and $r_{c}$.

\begin{figure}
   \centering
    \includegraphics[width=8cm]{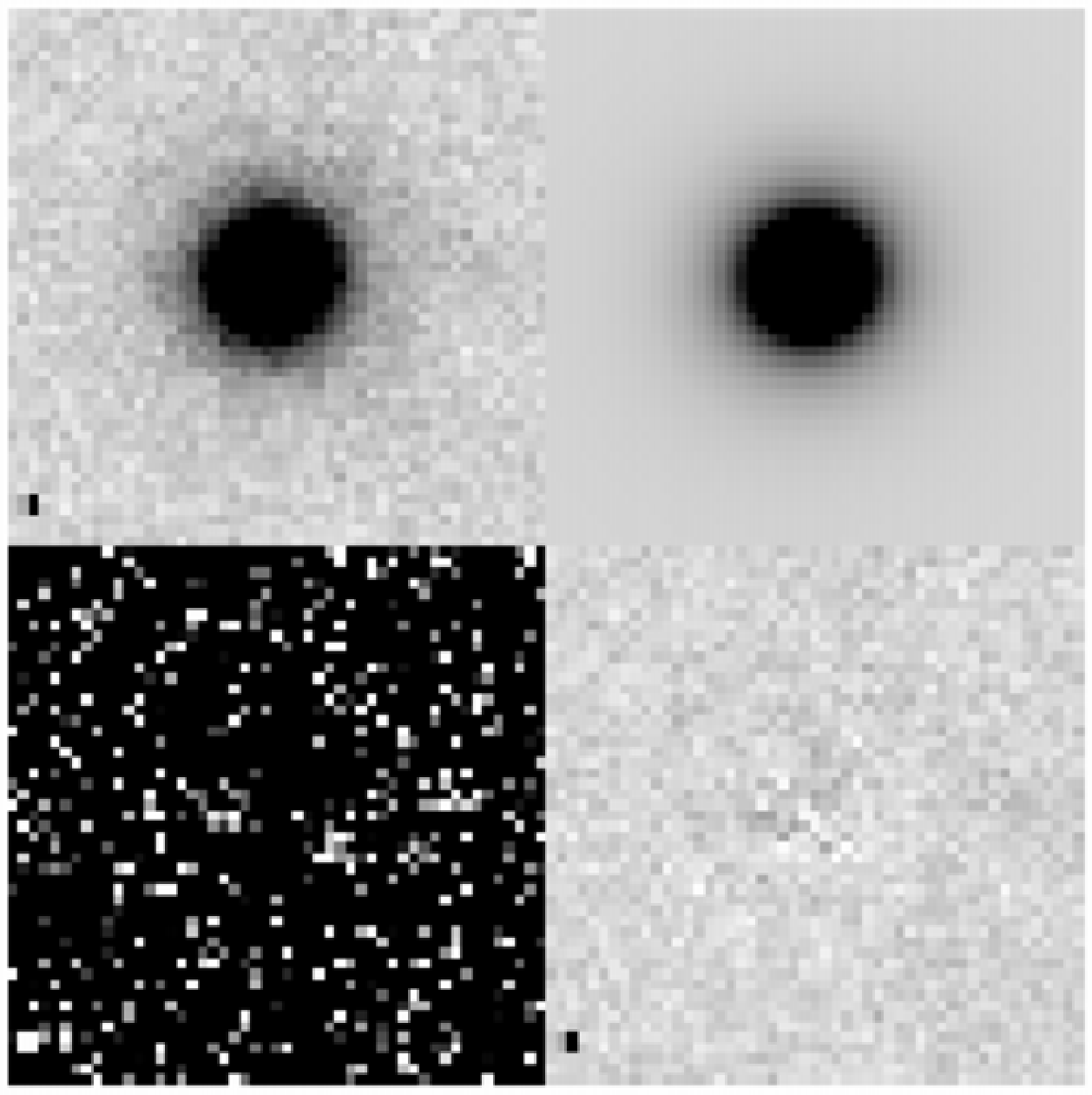}
      \caption{{\bf Upper left:} Original image of the cluster candidate HHG128. {\bf Upper right:} the model constructed 
using ishape and assuming a King profile, after the convolution with the PSF. {\bf Lower left:} The weights assigned by 
ishape to each pixel. {\bf Lower right:} The residuals after subtracting the original image with the model. Little 
structure is seen beyond the photon noise associated with the cluster's light.}
         \label{fig.ishape}
   \end{figure}

The fit radius was 25 pixels or $1.725\arcsec$, corresponding to 33 pc if a distance of 4 Mpc is assumed ($1\arcsec \sim19$ pc). 
Smaller radii were
tried, but the subtracted images showed noticeable residuals in the wings. With the sole exception of $c$, the derived
parameters do not change significantly with fitting radius. Fig.~\ref{fig.ishape} shows a typical example for one of
the known cluster candidates (HHG128).

Formally, the lowest $\chi^{2}$ is achieved with $c$ as a free parameter. However, as this is the most
uncertain of the parameters if left free, we have used fixed values of $c=30$ and $c=100$ to estimate the
actual errors using these plausible values for clusters in our Galaxy (Djorgovski \& Meylan~\cite{djorgovski93}).
Hence, all derived sizes, ellipticities and position angles (PA) are the average from the fits assuming King models
with $c=30,100$ and free. Their corresponding errors are the $\sigma$ in each case.

\begin{table*}
\caption{Results from the fitting procedure. The first column is the identification of the cluster. Then follow
the core radius $r_{c}$ in pc (assuming $d=4$ Mpc), ellipticity and the position angle (PA). Column~5 gives the 
concentration parameter $c~(=r_{t}/r_{c})$. 
The effective or half-light radius $r_{e}$ (in pc) is given in Column~6. Column~7 lists the projected galactocentric 
distance in kpc, using $\alpha=201.364995$ and $\delta=-43.01917$ as the center of \object{NGC\,5128}. 
The central surface brightness in 
$V$, $\mu_{0}(V)$ is listed in Column~8 (see text).}
\label{tab.results}
\centering
\begin{tabular}{l c c r c c l c c}     
\hline\hline
cluster ID	& $r_{c}$	 & e                 	& PA 	         	& c   &  $r_{e}$ 	& $R_{gc}$     & $\mu_{0}(V)$ \\
(HHG)	   	& (pc)	 	 &                   	& (deg.)  	        &     &  (pc)    	& (kpc) & (arcsec$^{-2}$)  \\
\hline

022 (WHH001)    &  0.73  $\pm$ 0.39  & 0.12  $\pm$ 0.00  &   2.  $\pm$  0.6 &  61.9   &  2.96 $\pm$ 1.58 & 14.2 &   13.85  \\
022 (WHH001)    &  0.75  $\pm$ 0.39  & 0.12  $\pm$ 0.01  &   1.  $\pm$  1.4 &  58.0   &  2.94 $\pm$ 1.53 & 14.2 &   13.88  \\
032     	&  1.25  $\pm$ 0.44  & 0.30  $\pm$ 0.00  & -21.  $\pm$  0.1 &  44.8   &  4.36 $\pm$ 1.53 & 29.0 &   15.23  \\
051 (WHH004)    &  0.41  $\pm$ 0.54  & 0.00  $\pm$ 0.01  &  53.  $\pm$ 13.7 & 317.5   &  3.73 $\pm$ 4.81 & 17.4 &   14.01  \\
080     	&  3.32  $\pm$ 0.32  & 0.08  $\pm$ 0.01  &  32.  $\pm$  4.6 &  36.4   & 10.42 $\pm$ 1.01 & 26.7 &   18.38  \\
086  (WHH031)   &  2.14  $\pm$ 0.37  & 0.24  $\pm$ 0.00  &  31.  $\pm$  0.5 &  35.5   &  6.62 $\pm$ 1.16 & 19.7 &   17.34  \\
093     	&  2.67  $\pm$ 0.29  & 0.18  $\pm$ 0.00  & -48.  $\pm$  0.1 &  43.1   &  9.08 $\pm$ 1.00 & 13.3 &   17.96  \\
102    		&  1.99  $\pm$ 2.95  & 0.38  $\pm$ 0.02  &  11.  $\pm$  0.7 &  30.0   &  5.69 $\pm$ 8.41 & 32.7 &   17.61  \\
104    		&  3.18  $\pm$ 0.25  & 0.11  $\pm$ 0.00  & -18.  $\pm$  1.0 &  21.4   &  7.71 $\pm$ 0.61 & 29.6 &   18.39  \\
106 (WHH029)   	&  0.63  $\pm$ 0.46  & 0.00  $\pm$ 0.01  & -28.  $\pm$ 22.6 &  62.9   &  2.58 $\pm$ 1.86 & 14.9 &   15.31  \\
128    		&  0.93  $\pm$ 0.31  & 0.07  $\pm$ 0.00  &  42.  $\pm$  2.8 &  30.9   &  2.69 $\pm$ 0.89 & 15.7 &   16.20  \\
129    		&  3.21  $\pm$ 0.20  & 0.28  $\pm$ 0.00  & -90.  $\pm$  0.9 &  75.0   & 14.33 $\pm$ 0.90 & 30.8 &   18.99  \\
141    		&  5.21  $\pm$ 0.21  & 0.06  $\pm$ 0.01  & -65.  $\pm$  8.1 &  46.5   & 18.42 $\pm$ 0.77 & 14.9 &   19.75  \\
210    		&  3.97  $\pm$ 0.37  & 0.06  $\pm$ 0.02  & -11.  $\pm$ 23.8 &  33.1   & 11.90 $\pm$ 1.12 & 33.2 &   20.06  \\
225    		&  5.40  $\pm$ 0.41  & 0.19  $\pm$ 0.02  & -17.  $\pm$  1.3 &  43.9   & 18.55 $\pm$ 1.42 & 30.5 &   20.73  \\
327    		&  0.71  $\pm$ 0.51  & 0.17  $\pm$ 0.04  & -64.  $\pm$  2.4 &  39.2   &  2.32 $\pm$ 1.65 & 13.1 &          \\
N1 (C40)     	&  1.38  $\pm$ 0.38  & 0.29  $\pm$ 0.00  &  79.8 $\pm$  0.4 &  40.3   &  4.54 $\pm$ 1.26 & 24.5 &   15.94  \\
N3      	&  2.19  $\pm$ 0.08  & 0.02  $\pm$ 0.01  &  55.7 $\pm$ 30.0 &  34.6   &  6.69 $\pm$ 0.25 & 15.1 &   19.28  \\
N3      	&  2.09  $\pm$ 0.16  & 0.06  $\pm$ 0.03  & -54.6 $\pm$ 10.4 &  50.8   &  7.71 $\pm$ 0.59 & 15.1 &   19.27  \\
N5      	&  1.62  $\pm$ 0.21  & 0.09  $\pm$ 0.02  & -23.6 $\pm$  9.7 &  22.4   &  4.00 $\pm$ 0.53 & 12.9 &   18.36  \\
N7 (PFF06)     	&  1.06  $\pm$ 0.44  & 0.36  $\pm$ 0.00  &  88.8 $\pm$  0.8 &  29.2   &  2.99 $\pm$ 1.25 & 15.7 &   15.45  \\
N8      	&  1.71  $\pm$ 0.59  & 0.38  $\pm$ 0.01  & -81.7 $\pm$  0.0 &  17.6   &  3.76 $\pm$ 1.30 & 15.1 &   17.64  \\
N9      	&  1.34  $\pm$ 0.31  & 0.00  $\pm$ 0.04  & -90.  $\pm$ 21.8 &  24.6   &  3.47 $\pm$ 0.80 & 14.2 &   19.21  \\
N10 (PFF09)    	&  1.13  $\pm$ 0.37  & 0.14  $\pm$ 0.01  &  85.3 $\pm$  0.9 &  44.8   &  3.92 $\pm$ 1.30 & 17.0 &   16.29  \\
N11     	&  1.20  $\pm$ 0.25  & 0.13  $\pm$ 0.01  &  23.2 $\pm$  2.8 &  34.1   &  3.64 $\pm$ 0.78 & 16.9 &   17.36  \\
N12     	&  1.25  $\pm$ 0.28  & 0.14  $\pm$ 0.02  & -25.8 $\pm$  1.2 &  22.9   &  3.14 $\pm$ 0.70 & 17.1 &   17.14  \\
N21     	&  3.32  $\pm$ 0.32  & 0.08  $\pm$ 0.01  &  31.7 $\pm$  4.6 &  36.4   & 10.42 $\pm$ 1.01 & 26.1 &   19.15  \\
N23     	&  2.61  $\pm$ 0.24  & 0.23  $\pm$ 0.00  &  47.9 $\pm$  0.2 &  43.2   &  8.89 $\pm$ 0.82 & 20.9 &   18.57  \\
N24     	&  2.40  $\pm$ 0.17  & 0.12  $\pm$ 0.03  & -73.  $\pm$ 20.2 &  56.8   &  9.37 $\pm$ 0.68 & 15.1 &   19.90  \\
N25     	&  2.98  $\pm$ 0.21  & 0.08  $\pm$ 0.02  &  44.  $\pm$ 13.7 &  29.1   &  8.38 $\pm$ 0.60 & 14.8 &   19.94  \\
N26 (PFF092)    &  0.98  $\pm$ 0.35  & 0.07  $\pm$ 0.01  &  44.5 $\pm$  4.2 &  55.8   &  3.79 $\pm$ 1.34 & 12.3 &   16.33  \\
N30     	&  2.27  $\pm$ 0.26  & 0.10  $\pm$ 0.03  &  11.2 $\pm$ 19.9 &  31.3   &  6.61 $\pm$ 0.78 & 24.5 &   18.20  \\
N32     	&  3.94  $\pm$ 0.32  & 0.12  $\pm$ 0.01  &  -8.1 $\pm$  3.1 &  51.4   & 14.61 $\pm$ 1.19 & 20.7 &   19.43  \\
N33     	&  0.99  $\pm$ 0.46  & 0.01  $\pm$ 0.03  &  -3.2 $\pm$ 26.4 &  24.0   &  2.53 $\pm$ 1.16 & 21.0 &   16.76  \\
N34     	&  4.39  $\pm$ 0.12  & 0.09  $\pm$ 0.00  & -23.4 $\pm$  9.1 &  51.0   & 16.24 $\pm$ 0.45 & 26.5 &   19.91  \\
N35     	&  3.65  $\pm$ 0.76  & 0.17  $\pm$ 0.03  & -77.3 $\pm$  4.2 &  71.1   & 15.84 $\pm$ 3.31 & 27.8 &   20.45  \\
N37     	&  4.10  $\pm$ 0.21  & 0.09  $\pm$ 0.04  &  52.6 $\pm$ 21.2 &  39.3   & 13.35 $\pm$ 0.70 & 25.1 &   20.91  \\
N41     	&  2.41  $\pm$ 0.16  & 0.24  $\pm$ 0.01  & -40.9 $\pm$  0.0 &  49.7   &  8.80 $\pm$ 0.60 & 31.9 &   18.80  \\
N42     	&  2.64  $\pm$ 0.20  & 0.09  $\pm$ 0.00  & -39.1 $\pm$  4.3 &  31.0   &  7.65 $\pm$ 0.58 & 30.7 &   17.35  \\
\hline
\end{tabular}
\end{table*}

Ishape does not return either the core radius $r_{c}$ or the tidal radius $r_{t}$. Instead, some assumptions must be made
to convert the given FWHM and concentration parameter into these more familiar quantities. A discussion of this can be found
in Larsen~(\cite{larsen01a}) and Larsen~(\cite{larsen04}), where both a relation between the FWHM and $r_{c}$ and a 
numerical approximation for the effective (half-light) radius $r_{e}$ are presented:

\begin{equation}
FWHM = 2 \left[ \biggl[ \sqrt{1/2} + \frac{1-\sqrt{1/2}}{\sqrt{1 + c^{2}}}\biggr]^{-2} -1 \right]^{1/2} r_{c}
\end{equation}

\begin{equation}
r_{e} / r_{c} \approx 0.547 c^{0.486}
\end{equation}

\noindent
The latter is good to $\pm$2\% when $c>4$. These relations are valid when the King profiles are circularly symmetric.
In our case (2-D fit) the average of the FWHM along the minor and major axis was used to compute $r_{c}$ and $r_{e}$.
The values are listed in Table~\ref{tab.results}.

The ellipticities (Column~3 in Table~\ref{tab.results}) are measured at the FWHM as $e=1 - b/a$, where $a$ and $b$ are
the semi-major and semi-minor axes respectively. Their values and comparison with other GCSs will be discussed in Sect.~4.1.

The central surface brightness of the clusters can be derived assuming that a King profile remains a good representation
of the light distribution of the cluster towards the center, even if this part cannot be resolved by our images.
The central surface brightness $\mu_{0}$ of a King profile and total luminosity $L(R)$ within a radius $R<r_{t}$
are related by (Larsen~\cite{larsen01a}):

\begin{equation}
\mu_{0} = k \left( 1 - \frac{1}{\sqrt{1 + c^{2}}} \right)^{2}
\end{equation}

\begin{equation}
L(R) = \pi k \left( r_c^2 \mathrm{ln} \left(1 + \frac{R^2}{r_c^2} \right) + \frac{R^2}{1+c^2} - \frac{4r_c^2}{\sqrt{1+c^2}}\left(\sqrt{1+\frac{R^2}{r_c^2}}-1\right)\right)
\end{equation}

Note that Eq.~(4) is essentially a definition of the magnitude zeropoint parameter $k$;
for the range of $c-$values of interest here, we have $k \simeq \mu_0$.
$R$ was set to 20 pixels ($1.38\arcsec$) and aperture photometry was consequently performed to compute the total luminosity $L(R)$
within this radius, which is roughly three times larger than the typical FWHM. As no Landolt standards were observed 
(since the nights were not photometric), we have used the Washington photometry from Harris~(\cite{harris04a}), which 
includes virtually everything in the field of \object{NGC\,5128} in order to calibrate our photometry. For this
purpose an aperture correction between their and our aperture radius had to be computed. The curves of growth (i.e. the $R$ 
magnitude vs. the aperture radius) show that this correction is $\Delta R = -0.130 \pm 0.005$. Our $R$ photometry can be
directly compared to the $T_{1}$ filter as the difference is $\sim 0.01$ for 
old, globular cluster-like objects (see Geisler~\cite{geisler96a}).
In addition to the aperture correction, we used $E_{B-V} = 0.11$ (Schlegel et~al.~\cite{schlegel98}) 
to correct for galactic absorption. Since all of our clusters are well away from the 
central dust lane, the reddening should be uniform.

The central surface brightness in the $R$ band for each cluster is thus derived using eqs. (4) and (5). However, most values 
found in the literature are given only for the $V$ filter. Hence, we have taken an average colour of $V-R=0.5$ to estimate
$\mu_0(V)$. The results are listed in Column~8 of Table~\ref{tab.results}. The spread in the intrinsic colour remains
the main uncertainty for the values listed.

Two clusters (HHG022 and N3, see Tables 1-3) were observed twice under similar conditions. They allowed us to test the 
internal accuracy of the derived solutions. As can be seen from these Tables, the results are encouraging given the very 
good agreement between the listed values. Furthermore, a much more telling  -and fully independent- comparison is that with
Harris et~al.~(\cite{harris02}). Their typical sizes and ellipticities are comparable to those derived by us.

\section{Discussion}

\subsection{Ellipticity}

The Galactic GCs show, on average, little elongation, apart from a few cases (e.g. $\omega$~Cen with $e=0.19$, 
Frenk \& Fall~\cite{frenk82}). The range of ellipticities of GCs is interesting to study 
given that systematic differences have been found for other systems in the Local Group. Geisler \& 
Hodge~(\cite{geisler80}) concluded that only a few massive (especially intermediate age) clusters 
in the LMC are round, the large majority showing significant 
ellipticities. A similar feature was observed for GCs in M31 (Barmby et.~al.~\cite{barmby02}), 
where the brightest member G1 (=Mayall II) has $e=0.25\pm0.02$ measured from HST images 
(Rich et~al.~\cite{rich96}). A common property of all these 
very elongated clusters is their high luminosity. In fact, the brightest GCs in our Galaxy, as well as
in M31 and in the LMC, are the most flattened in their respective galaxies (van~den~Bergh~\cite{vandenbergh84}).
Hesser et~al.~(\cite{hesser84}) also noted that 4 of the 6 brightest known \object{NGC\,5128} GCs
were noticeably flattened.

\begin{figure}
   \centering
    \includegraphics[width=8cm]{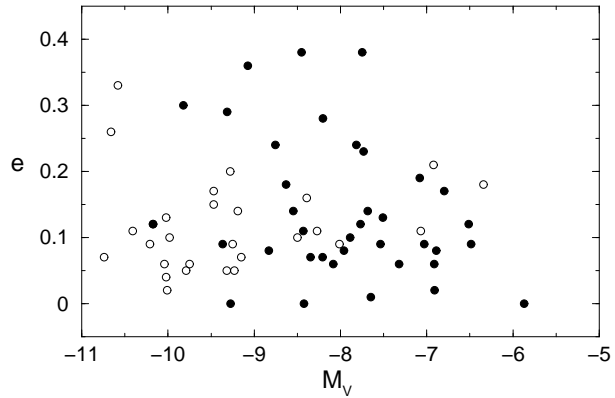}
      \caption{The ellipticity $(e=1-b/a)$ as a function of the cluster's luminosity in the $V$ band, $M_{V}$.
The same behaviour is observed for Galactic and M31 clusters, in the sense that brighter clusters tend 
to be more flattened and there are no relatively faint, elliptical clusters. Our data are indicated 
by filled circles, while open circles represent the clusters from Harris et~al.(~\cite{harris02}).}
         \label{fig.MV_ellip}
   \end{figure}

Although our sample is still small, we do observe a similar tendency of bright clusters to have
higher ellipticities, as shown in Fig.~\ref{fig.MV_ellip}. It is intriguing to suppose that
such clusters may be the stripped nuclei of former nucleated dwarf elliptical galaxies
(e.g. Martini \& Ho~\cite{martini04}), as is often supposed for $\omega$~Cen. However,
note that the second most luminous Galactic GC (M54) is nearly spherical, although it is now
generally regarded as the former nucleus of the Sgr dSph (e.g. Layden \& Sarajedini~\cite{layden00}).

\begin{figure}
   \centering
    \includegraphics[width=8cm]{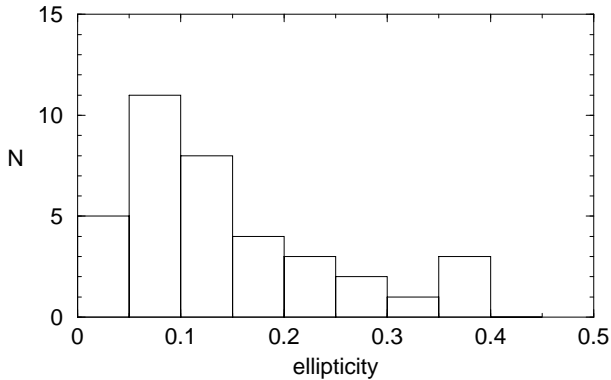}
      \caption{Histogram of the ellipticities. The reader is referred to Fig.~7 of Harris et~al.~(\cite{harris02}) 
for a similar plot for other GCSs. The strikingly high ellipticities observed in a few cases in our sample put them among 
the most flattened GCs observed so far.}
         \label{fig.histo_ellip}
   \end{figure}

The histogram of our ellipticities (Fig.~\ref{fig.histo_ellip}) is in very good agreement with 
the samples studied by Barmby et~al.~(\cite{barmby02}) for M31 GCs and Holland 
et~al.~(\cite{holland99}) and Harris et~al.~(\cite{harris02}) for Cen A GCs, although it goes 
to somewhat larger values. 
The bump in the interval $e=0.35-0.4$ is not present in any of the samples described above, 
although H2002 find a cluster with $e=0.33$. Objects having this extreme elongation should not 
necessarily be discarded as GCs. Indeed, the H2002 object is certainly a cluster. In addition,
Larsen~(\cite{larsen01a}), using HST images, studied a bright
cluster in NGC\,1023 and estimated its ellipticity as $0.37\pm0.01$, thus being one of the most 
flattened GCs known so far. On the other hand, we must recognize from the above discussion that 
galaxies certainly must contaminate our sample and these may well help to skew the distribution 
to higher ellipticity (see Section~5 for a discussion on the background contamination).

\subsection{Size, luminosity and surface brightness}

\begin{figure}
   \centering
    \includegraphics[width=8cm]{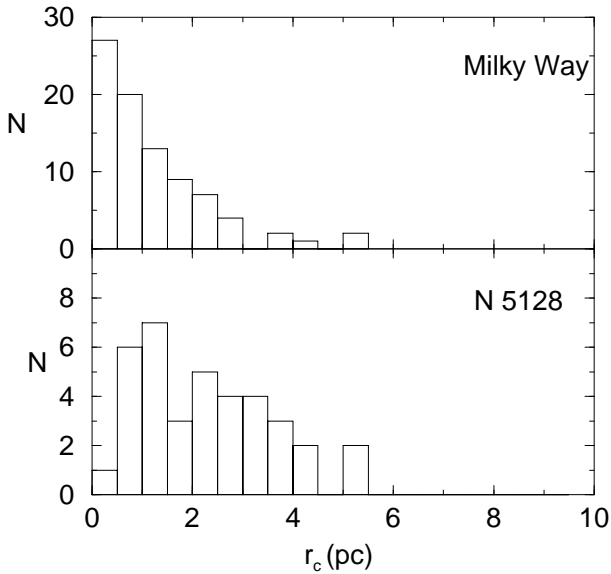}
      \caption{Histogram of the core radius $r_{c}$ for Galactic clusters (top panel, 85 objects) and \object{NGC\,5128} (bottom panel, 39 objects).}
         \label{fig.rc_comp}
   \end{figure}

Both the effective radii $r_e$ and core radii $r_c$ are listed in Table~3. The core radii agree very well with the 
range of values for GCs in NGC 5128 given in 
Harris et~al.~(\cite{harris02}), and as can be seen from Fig.~\ref{fig.rc_comp}, are also well in the range of Galactic GCs. 
The dip observed at $r_c < 0.5$ pc is probably due to a selection effect and is also present in the H2002 sample,
i.e. the known GCs tend to be the largest, most easily resolved.

\begin{figure}
   \centering
    \includegraphics[width=8cm]{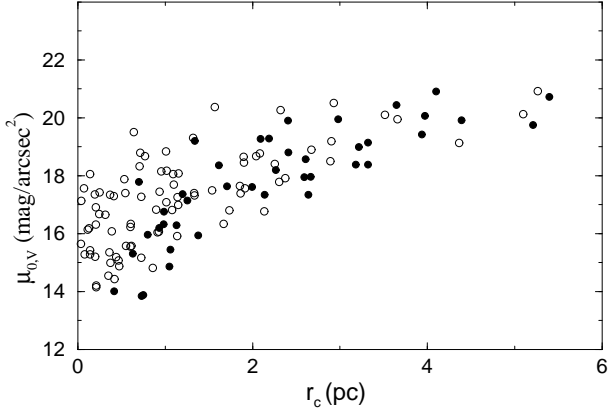}
      \caption{Comparison of the central surface brightness in the V-band $\mu_{0}(V)$ 
vs. $r_c$ for
clusters in \object{NGC\,5128} (filled circles) and Galactic GCs (open circles).}
   \label{fig.rc_muV}
   \end{figure}

Fig.~\ref{fig.rc_muV} compares the core radii and the central surface brightness $\mu_0(V)$ between clusters of 
\object{NGC\,5128} (filled circles) and Galactic ones (open circles). No major systematic difference is observed and both groups show
a similar spread in their central surface brightness at a given $r_c$. 
However, the tendency of the smaller clusters in \object{NGC\,5128} to have higher central
surface brightness could indicate a selection bias. This is very plausible as our detection is based on the presence of
residuals and faint compact clusters would be hardly noticeable.

Although the core radii are not surprising, the effective radii are generally somewhat larger than the 
typical 2-7 pc observed for MW GCs, with a few extreme clusters with $r_e \sim$15 pc. 
However, clusters with sizes of $\sim$30 pc have been observed in M31
(Huxor et~al.~\cite{huxor05}) at galactocentric distances between 15 and 35 kpc.
Again, we are certainly biassed against the smaller, more compact clusters.

In addition to these large GCs, Martini \& Ho~(\cite{martini04}) have derived masses and sizes for a sample of 14 bright clusters in \object{NGC\,5128}.
Their masses are in the range $10^6 - 10^7 M_{\odot}$ and thus some clusters are even more massive than some dwarf galaxies.
In our Galaxy, the most massive GC ($\omega$~Cen) may well be a stripped dwarf nucleus (see Martini \& Ho~\cite{martini04}
and references therein). A similar origin has been proposed for G1 in M31 (e.g. Bekki \& Freeman~\cite{bekki03}, 
Bekki \& Chiba~\cite{bekki04}). By implication, these very massive \object{NGC\,5128} GCs may also have had such an origin.
On the other hand, de~Propris et~al.~(\cite{depropris05}) have compared surface brightness profiles
of the nuclei of Virgo dwarf galaxies with those of UCDs. They concluded that UCDs are more extended and brighter
than the dwarf nuclei, so the ``threshing scenario'' is unlikely.

\begin{figure*}
   \centering
    \includegraphics[width=13.8cm]{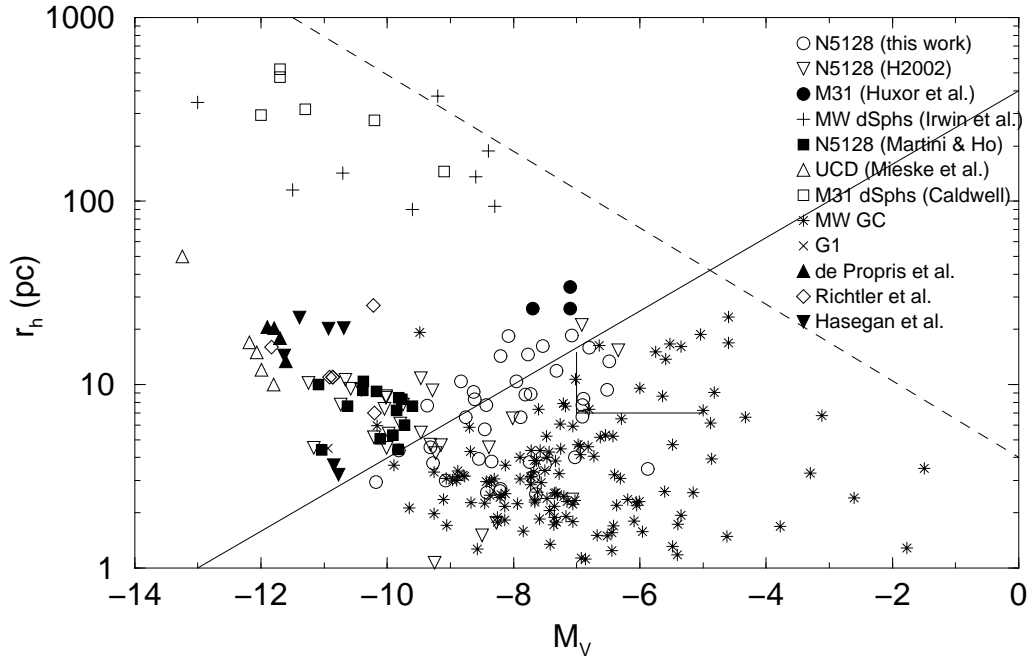}
      \caption{Half-light radius $r_{h}$ (in pc) versus $M_{V}$. A distance of 4 Mpc for 
\object{NGC\,5128} has been assumed. The references for the different symbols follow. Open circles 
are the GCs analysed in this work. Filled circles are the luminous clusters found in the 
halo of M31 by Huxor et~al.~(\cite{huxor05}). Downward-pointing empty triangles represent the data 
for GCs in \object{NGC\,5128} using HST images from Harris et~al.~(\cite{harris02}). The plus signs 
are dwarf Spheroidals associated with the Milky Way (Irwin \& Hatzidimitriou~\cite{irwin95}). 
The sample of massive clusters in \object{NGC\,5128} studied by Martini \& Ho (\cite{martini04}) are plotted 
using filled squares. Open squares show the dwarf Spheroidals associated with M31 (Caldwell 
et~al.~\cite{caldwell92} and Caldwell~\cite{caldwell99}). 
UCDs from Mieske et~al.~(\cite{mieske02}) are represented by the upward-pointing empty triangles. 
The MW GCs are the asterisks (data taken from the catalog of Harris~\cite{harris96}). 
G1 (=Mayall II), the brightest GC in M31, is shown with a cross. The data from de~Propris
et~al.~(\cite{depropris05}) and Ha\c{s}egan et~al.~(\cite{hasegan05}) are shown with the 
filled upward and downward pointing triangles, respectively. Finally, the empty diamonds
represent the striking clusters/UCDs found by Richtler et~al.~(\cite{richtler05}) around NGC\,1399.
The equation $log r_h = 0.2 M_V + 2.6$ from van~den~Bergh \& Mackey~(\cite{vandenbergh04})
is the solid line and the dashed line shows a value of constant average surface brightness 
with $r_h$. The solid L-shape indicates the region (above and to the right) where FFs are 
found. GCs form a continuum in this diagram and even approach the region occupied by UCDs.}
   \label{fig.MV_rh}
   \end{figure*}

Whatever the physical processes that led to the formation of these very
massive clusters, their location in $r_h - M_V$ parameter space
is quite different from what is observed for MW GCs, as is clear in Fig.~\ref{fig.MV_rh}. 
The solid line shows the equation derived by van~den~Bergh \& 
Mackey~(\cite{vandenbergh04}) who found that only 2 MW GCs lie above this line:
$\omega$~Cen and NGC\,2419.
The data have been primarily taken from Huxor et~al.~(\cite{huxor05}), to which
we have added the samples of de~Propris et~al.~(\cite{depropris05}), Richtler et~al.~(\cite{richtler05})
and Ha\c{s}egan et~al.~(\cite{hasegan05}). The references for each dataset and their 
symbols are given in Fig.~\ref{fig.MV_rh}.

In spite of the difference between the most massive NGC 5128 GCs and
their MW counterparts, the most striking feature of this diagram is that 
{\em there are essentially no longer any gaps in the distribution of the ensemble of GCs.\/}
When Huxor et al. first presented this diagram, they used it to point to
the unique position of their newly-discovered M31 clusters. However, 
the addition of our data and that of H2002 now nicely fills the gaps that
were present in  the Huxor et~al.~(\cite{huxor05}) version. Many \object{NGC\,5128} 
GCs are found above the van~den~Bergh \& Mackey~(\cite{vandenbergh04}) line. 
Clearly, this line no longer appears to have any special significance.
In particular, we find a number of \object{NGC\,5128} clusters that are large and of 
intermediate luminosity, falling only slightly below the Huxor et~al. M31 clusters. 
In addition, several \object{NGC\,5128} clusters are found in the region 
formerly inhabited almost exclusively by the Faint Fuzzies of Larsen \& Brodie~(\cite{larsen00}). 
Note that the majority of the \object{NGC\,5128} clusters in this region are 
also red and presumably metal-rich, as are the Faint Fuzzies. The 3 M31 clusters, 
on the other hand, are rather blue.
Therefore, large, faint clusters are not exclusive to lenticular galaxies.
However, note that FFs are also distinguished by their disk-like kinematics.
This cannot be further studied in our sample without spectroscopic observations 
and we are in the process of obtaining such data.

Thus, our data serve to illustrate that, although they 
exhibit a range of 1-2 orders of magnitude in both luminosity and size,
{\em GCs can not be broken down into well-separated, distinct subtypes but instead
form a continuum\/}.  This is in keeping with the assessment of Larsen~(\cite{larsen02b}). 
This continuum now extends to the realm of the Ultra Compact Dwarfs (e.g. Hilker 
et~al.~\cite{hilker99}, Drinkwater et~al.~\cite{drinkwater02}), lending support to the idea 
that these objects are indeed similar (Martini \& Ho~\cite{martini04}) and
have similar origins (Mieske et~al.~\cite{mieske02}). Note that Ha\c{s}egan
et~al.~(\cite{hasegan05}) find several objects in their Virgo Cluster ACS survey 
that they term ``dwarf-globular transition objects'' (DGTOs). Several of these fall close 
to the main locus of GCs in Fig.~\ref{fig.MV_rh} and several lie closer to that 
of previously identified UCDs. The existence of such objects further serves to 
fill in the parameter space in this diagram. Note, however, that Ha\c{s}egan 
et~al.~(\cite{hasegan05}) suggest that at least UCDs and GCs may be best 
distinguished via other parameters, e.g. M/L ratio.

\subsection{Correlations with galactocentric distance}

Several of the structural parameters may depend on the galactocentric distance $R_{gc}$, including the size,
brightness and concentration. For the Milky Way system, van~den~Bergh et~al.~(\cite{vandenbergh91}) have noticed that 
$r_h$ is correlated with $R_{gc}$, with larger clusters observed at larger distances.

A similar correlation was observed for the GCS of M31 (Barmby et~al.~\cite{barmby02}), but it is not so clear for 
the H2002 sample. This is in part, because of the different cameras used for inner and outer clusters and, most 
important, because of projection effects over a small sample, which can blur any subtle trend with the actual, 
three-dimensional galactocentric distance.

\begin{figure}
   \centering
    \includegraphics[width=8cm]{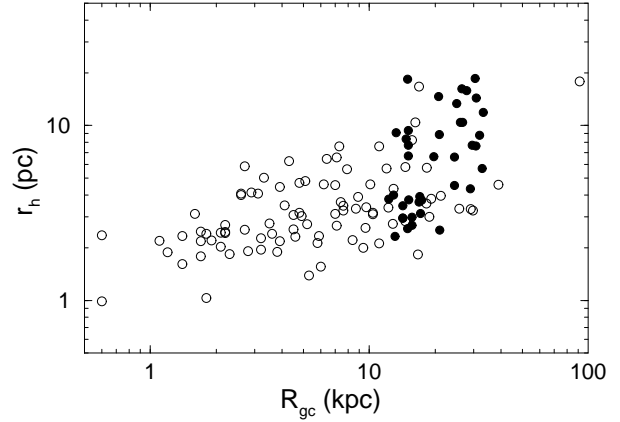}
      \caption{Dependence of the half-light radius $r_{h}$ on   the projected distance of the clusters to
the optical center of \object{NGC\,5128} (filled circles). A distance of 4 Mpc has been assumed for the galaxy. For
comparison, the open circles represent Milky Way GC, taken from the catalog of Harris~(\cite{harris96}).}
         \label{fig.sep_rh}
   \end{figure}

Our data show, nevertheless, a clear agreement with the trend defined by the Galactic GCs (see Fig.~\ref{fig.sep_rh}).
On the logarithmic scale, it is apparent that the observed range of $r_h$ in \object{NGC\,5128} matches that in the Milky Way,
although our clusters have on average larger projected distances (see also Fig.~\ref{fig.n5128_dss}). 
More clusters at smaller distances need to be observed
before a better comparison can be made. Regarding the luminosity, we do not see any correlation with $R_{gc}$.
In analogy to Fig.10 of H2002, we searched for possible correlations between the concentration parameter and
ellipticity with $R_{gc}$, but none were observed.

\subsection{Metallicity and size}

To estimate the metallicity of the clusters, we have used the relation from Harris \& 
Harris~(\cite{harris02b}) between the Washington system $(C-T_1)$ colour and metallicity,
and assumed a uniform reddening of $E(B-V)$=0.11 (Schlegel et~al.~\cite{schlegel98}) 
for all clusters. Given the very large distances of our clusters from 
the central dust lane, this assumption should be valid.
No trend is seen between the derived metallicities and $R_{gc}$, but due to the still small
sample, we do not further comment on this. Instead, we have arbitrarily divided the sample into metal-poor and
metal-rich at Fe/H=--1 (see Harris et~al.~\cite{harris04a}) and looked for systematic differences in the cluster size. 
The analysis performed on a number of galaxies ( see Kundu et~al.~\cite{kundu99}, Kundu \& Whitmore~\cite{kundu01}, 
Larsen et~al.~\cite{larsen01b}, Jord\'an~\cite{jordan04},
Jord\'an et~al.~\cite{jordan05}) indicate that metal--poor (blue) 
clusters have mean half-light radii  $\sim20\%$ larger 
than those of metal--rich (red) ones. However, this is {\em not\/} the case for the H2002 sample. Using the 
same metallicity cut, they find mean radii of $r_{h}=7.37 \pm 1.03$ and $r_{h}=7.14 \pm 0.76$ for the 
metal--poor and metal--rich groups respectively, i.e. indistinguishable.

\begin{figure}
   \centering
    \includegraphics[width=8cm]{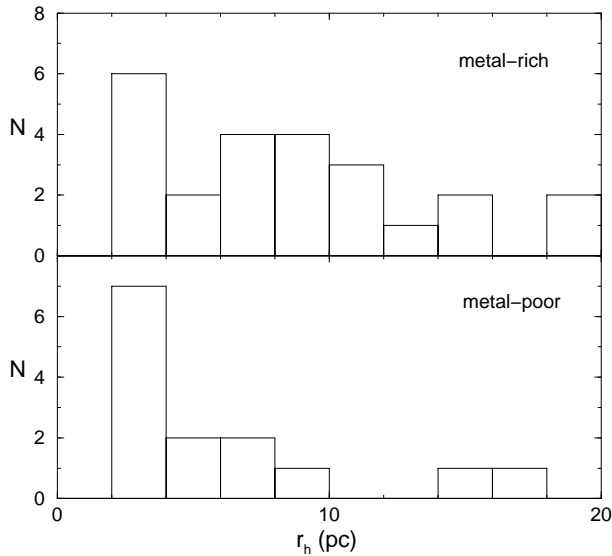}
      \caption{Half-light radii $r_h$ for the metal--poor and metal--rich subgroups. The metallicity is estimated
from the Washington C--T1 colour according to Harris \& Harris (2002) and the subgroups are divided at [Fe/H]=--1.}
         \label{fig.rc_met}
   \end{figure}

Our results are in better agreement with those of H2002 than with other studies. Fig.~\ref{fig.rc_met} 
shows the histogram of half-light radii $r_h$ for both subpopulations and it is clear that, if a 
relation between metallicity and cluster size exists, then it is opposite to that 
reported by Jord\'an~(\cite{jordan04}) and most previous WFPC2 studies of barely resolved clusters.
The returned median values are $r_{h}=4.1 \pm 1.2$ and $r_{h}=8.0 \pm 0.9$ pc
for the metal--poor and metal--rich
subgroups respectively. The paucity of large metal--poor clusters is apparent. 
It is striking that the Barmby et~al.~(\cite{barmby02}) histogram of $r_h$ 
for M31 GCs is broader for the metal--poor than for the metal--rich sample, which is contrary 
to what we see. This holds also if metallicities via broad-band colours are included. 

The general agreement of our results with those of H2002 lends credence to our
findings. Although the combined sample size is small, these data are
the best available in terms of number of pixels per resolved image and the
results should be the most robust. The possibility exists that some background galaxies
still contaminate our sample and only a few of them would be needed to affect the shape of the
size histograms. However, galaxy contamination should only be a small effect (see Section~5).
In fact, contaminating galaxies from the Woodley et~al.~(\cite{woodley05}) sample 
are preferentially in the blue (metal-poor) part
of our GC colour selection. This is opposite to the effect required 
to explain the size difference between metal-rich and metal-poor 
clusters by galaxy contamination.
Another possibility is that clusters in \object{NGC\,5128} are different from those observed in
other galaxies. Clearly, more observations are required to resolve this intriguing issue.

\subsection{Total Cluster Population}

Harris et~al.~(\cite{harris04b}) found that the \object{NGC\,5128} GCs follow a steep projected
radial distribution, of the form $\sigma \sim r^{-2}$, and that the clusters 
were mostly confined to the inner $\sim 10\arcmin$. However, note that {\em all\/} of 
our fields are {\em beyond\/} this radius, extending from $11\arcmin$
out to $29\arcmin$, with most outside $15\arcmin$. 
In the 30 fields we observed, we confirmed the GC nature of 1/3 of the 44 original 
Harris et~al.~(\cite{harris04a}) candidates
and found 22 other GC candidates, roughly 1 per field. The success of 
our GC search in  these distant regions of the galaxy's halo leads us to believe 
both that the radial extent of the \object{NGC\,5128} GC population is indeed large, 
with a significant population out to $30\arcmin$ and beyond, and that the number 
of clusters derived by Harris et~al.~(\cite{harris04a}) may be an underestimate.
A crude calculation derives a mean GC surface density for our sample that
agrees well with the value derived by Harris et~al.~(\cite{harris04b}), at the
mean galactocentric distance of our sample. However, our sample only includes 
a few of the {\em known} GCs at this radius. In addition, for their total population
calculation, Harris et~al.~(\cite{harris04b}) assumed a value derived from
the halo {\em light} profile, which fits well their GC profile out to 
$\sim13\arcmin$ but beyond this their GC densities appear higher than those of the
galaxy itself, as is generally observed.
If their surface density of clusters beyond $\sim 10\arcmin$ is only slightly
underestimated, this could lead to a significant increase in their
total GC population estimate of $\sim 1000$ clusters and this indeed 
appears to be the case.

Clearly, the final census of \object{NGC\,5128} GCs will not be achieved until a complete 
search of the outer halo to at least $45\arcmin$ is conducted. Radial velocity 
surveys would be very inefficient and time-consuming given the huge ($\sim$100 to 1) 
ratio of field interlopers to genuine GCs.
For optimum efficiency, such a search requires 
both the widest field coverage combined with very high spatial resolution
in order to resolve the clusters. Such an optimum combination is provided
by the IMACS instrument at Magellan and we are in the process of obtaining 
these data.

\section{Background Contamination}

Certainly, even after the selection criteria applied to our cluster candidates,
a (hopefully small) fraction of our objects consists of background galaxies.
Foreground stars can be surely discarded as part of the contaminating population,
because they cannot be observed as extended objects in our frames. The purpose of this
section is to give an idea of the contamination level and its effects on our results.

\subsection{Effects of including galaxies in our sample}

To see how galaxies behave in our diagrams (i.e. Fig~\ref{fig.rc_muV} and 
Fig.~\ref{fig.MV_rh}), we have re-analysed our frames and selected objects which are
either spectroscopically confirmed background galaxies or fall without doubt into this category
after visual inspection. All of the latter are well resolved objects, at least twice
as large as the clusters and many had substructures like disks or spiral arms.
Fainter objects were included only if they were spectroscopically confirmed galaxies. 
This is the case for WHH094, WHH127 and WHH43, all from Woodley et~al.~(\cite{woodley05}).

Our images of these galaxies were analysed in exactly the same way as with the cluster
candidates, i.e. light profiles were fitted via {\it ishape\/} and using the same stellar
PSF for the corresponding frame. The fitting radius and all parameters were handled
as was done for the cluster candidates, the only difference is that we are now working with
a sample of purely galaxies, so `normal' Washington colours and magnitudes and reasonable
fits with King profiles (i.e. little residual in the subtracted image) are not expected.

The derived structural parameters were then combined with the photometry from Harris
et~al.~(\cite{harris04a}).
After correcting for aperture size, reddening and forcing the galaxies to be at the
assumed distance for \object{NGC\,5128} (4 Mpc), we derived their surface brightnesses (a quantity
which is anyway independent of the distance), absolute magnitudes, half-light and core radii in
parsecs, etc.

\begin{figure}
   \centering
    \includegraphics[width=8.8cm]{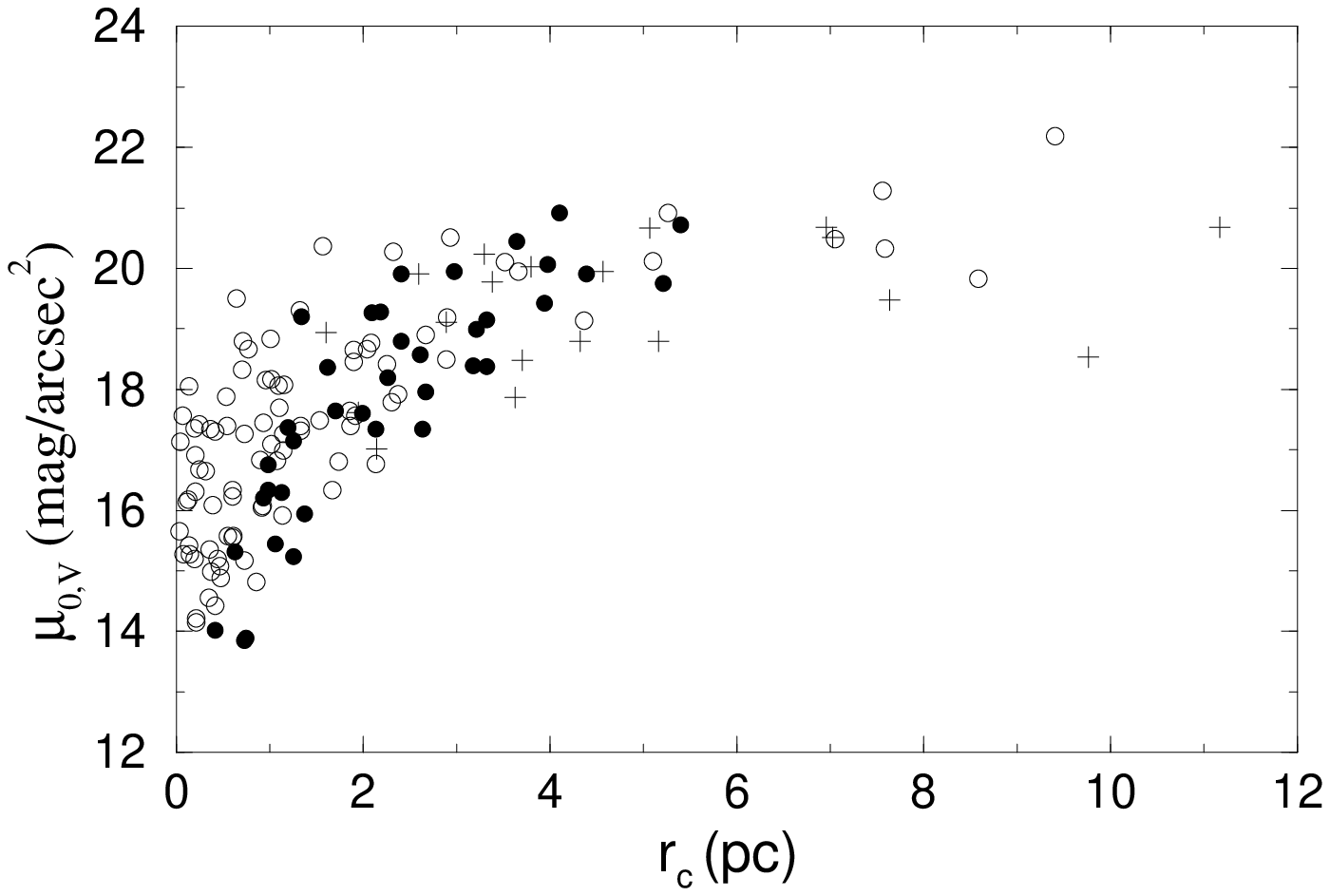}
    \includegraphics[width=8.8cm]{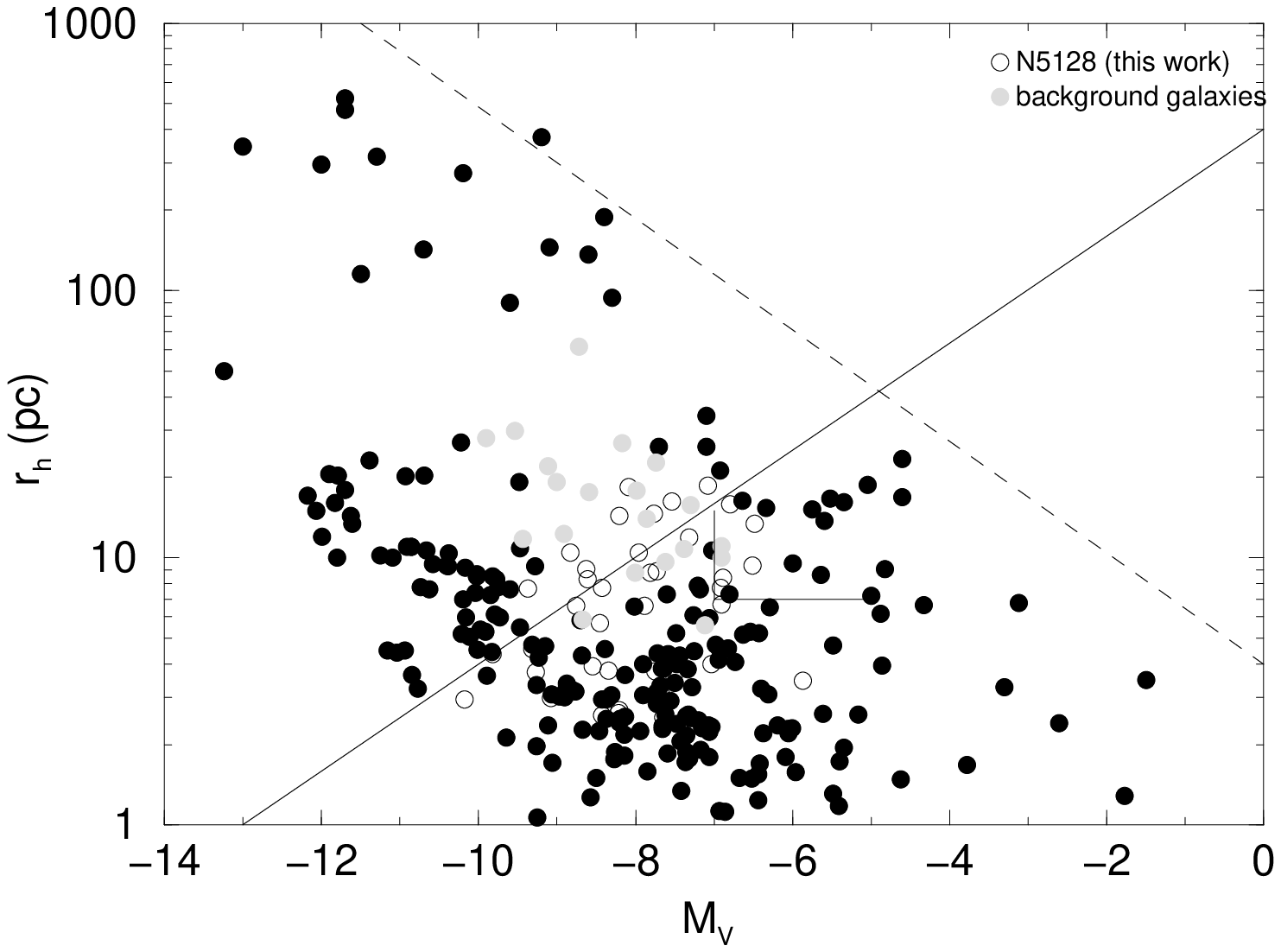}
      \caption{{\bf Upper Panel:} Same as Fig.~7 with background galaxies represented by the plus sign.
Besides the tendency to have larger $r_{c}$ and perhaps lower surface brightnesses than the
\object{NGC\,5128} (filled circles) and MW (open circles) GCs, galaxies cannot be clearly 
distinguished in this space.{\bf Lower Panel:} Same as Fig~8, but the analysed sample of galaxies
has been included and is represented by the grey circles. Open circles are the GC candidates and
filled circles represent all objects from the references given in Fig.~8.}
         \label{fig.background}
   \end{figure}

Strikingly, the King profile did in fact fit quite well to the majority of galaxies for which
no substructure was visible, at least within the inner 25 pixels. For disky or spiral galaxies,
the residuals were clear. Moreover, the galaxies occupy a similar region as the GCs in the
$\mu_{0}- r_{c}$ parameter space, as Fig.~\ref{fig.background} shows 
(upper panel), with perhaps a reasonable tendency to larger sizes and lower surface 
brightnesses. We stress that, in many aspects, the comparison between galaxies and clusters 
based on their structural parameters has no physical meaning. In this case, the derivation 
of $r_{c}$ via a King profile is not more than an artifact but it is adequate for the 
purpose of assessing the contamination level and to illustrate its behaviour.

Similarly, there is little observed difference in the $r_{h} - M_{V}$ space
(Fig.~\ref{fig.background}, lower panel). Indeed, the galaxies extend to higher sizes and span a region 
which is a bit above `normal' GCs in \object{NGC\,5128} and in our Galaxy in this logarithmic scale. 
Note, however, that the smallest galaxies in our sample
(again, small in the sense of $r_h$ obtained by brute force) is comparable to the mid-size/large
clusters. Truly large galaxies have no counterparts in our sample of GCs.

\subsection{The level of contamination}

To assess the level at which our diagrams and results could be affected by contamination
is the aim of this section.

Without doubt, the most reliable way to determine if a cluster candidate is actually a
background galaxy is via spectroscopy. Nevertheless, an independent image-based analysis
can also be used as a reasonable first measurement. For this, we made use of images
taken independently with the Gemini 8m Telescope and the GMOS Camera (Harris et~al.~2006).
These are of poorer seeing than the MagIC frames (0.7''--0.8'') and cover the central 
region of \object{NGC\,5128}, so there are no common objects. However, the published spectroscopic work
concentrates almost solely in this region and therefore, dozens of confirmed clusters and 
galaxies are available for study.

We apply the same reduction, analysis and selection criteria to the cluster-like
candidates within these GMOS fields and look for existing velocities for this sample.
Of course, this is only meaningful if one does not know {\it a priori\/} which objects are
the clusters. Thus, the correlation with spectra was done as the very last step.

First, the stellar PSF was made from 20-30 stars per frame. Then, this PSF was subtracted
from all objects and GC candidates were chosen among those which left a `doughnut' shaped
residual, exactly as was done for our original sample.
After this, we looked for their Washington photometry in H2004 and applied
the same colour and magnitude cuts. We then ran {\it ishape\/} again with the same parameters
as before and in exactly the same way.

All in all, we obtained 87 cluster-like objects for which structural parameters could be derived
and which passed the selection criteria. 42 of these had been previously observed
spectroscopically (Peng et~al.~\cite{peng04} and Woodley et~al.~\cite{woodley05}). 39 of
the 42 GC candidates turned out to be clusters and 3 are actually background galaxies.
There are, in addition, several GCs which we failed to detect, presumably due to the low
resolution of the images and their compactness.

This shows that the galaxy contamination level is roughly 7\% for the GMOS images and 
can be even lower for the MagIC study, which benefitted from much better resolution.

On the one hand, we have shown that background galaxies cannot be easily told from the clusters
based on their structural parameters and location in the diagrams, so that the risk exists
if no careful selection is performed, although in this case one still should not be
led to erroneous conclusions since the properties are similar. 
On the other hand, an independent test shows that this contamination is {\it at most\/} 7\% when 
the morphology and Washington colours are used as selection criteria. Thus, our results 
reflect by far the behaviour of GCs and {\it not\/} background galaxies. Note that 
the ratio of galaxies to GCs certainly increases with galactocentric distance, due to
the strong central concentration of the latter, and thus one might naively expect 
that the contamination level of our sample may be higher since it is more distant than
the GMOS fields. However, we have shown that, given sufficiently good seeing and our 
analysis techniques, one should be able to maintain the contamination to a low level
irrespective of galactocentric distance.

As explained in Section 2.3, Woodley et~al.~(\cite{woodley05}) find that two of our
cluster candidates are actually background galaxies. For one of them (HHG94),
our much better seeing shows that it is actually a cluster candidate
projected $1.6\arcsec$ from a galaxy observed. This separation is close to
the typical seeing for the Harris et~al.~(\cite{harris04a}) database, which was the
basis of the Woodley et~al.~(\cite{woodley05}) work. It is therefore reasonable
that the velocity obtained corresponds actually to that of the galaxy,
or a composite of the two velocities.

If this is indeed the case, then there is only one galaxy in the common sample
of 6 cluster candidates, which is roughly what we should expect from our 
results, given the small numbers involved. Our GMOS test is more 
definitive, so we give more weight to its result and estimate our final
galaxy contamination level at about 10\%.

\section{Conclusions}

We have obtained very high spatial resolution images in excellent seeing
conditions with the Magellan telescope + MagIC camera of 44 GC candidates in 
the outer regions of \object{NGC\,5128} from the list provided by Harris et~al.~(\cite{harris04a}). 
These data not only allow us to determine the nature of these candidates via spatial 
resolution but also allow us to derive their detailed structural parameters. 
This is the first time such parameters are determined for GCs beyond the 
Local Group from ground-based images. About one third of the candidates appear to be
bonafide GCs. In addition, we serendipitously 
discovered 18 new GC candidates and also derived their structural parameters.

We compare our cluster sample in detail with those of other GC samples in other 
galaxies with similar information available. We find that, in general, our 
clusters are similar in size, ellipticity, core radius and central surface 
brightness to their counterparts in other galaxies, in particular those in 
\object{NGC\,5128} observed with HST by Harris et~al.~(\cite{harris02}).
However, our clusters extend to higher ellipticities and larger half-light
radii than their Galactic counterparts, confirming the results of H2002.
Combining our results with those of Harris et~al.~(\cite{harris02}) fills in the gaps 
previously existing in the $r_h - M_V$ parameter space and indicates that any 
substantial difference between presumed distinct cluster types in this diagram, 
including for example the Faint Fuzzies of Larsen \& Brodie~(\cite{larsen00})
and the `extended, luminous' M31 clusters of Huxor et~al.~(\cite{huxor05})
is now removed and that clusters form a continuum. Indeed, this continuum now 
extends to the realm of the Ultra Compact Dwarfs.
The metal-rich clusters in our sample have half-light radii in
the mean that are almost twice as large as their metal-poor counterparts,
at odds with the generally accepted trend.
Finally, our discovery of a substantial number of new cluster candidates
in the relatively distant regions of the \object{NGC\,5128} halo suggest that current
values of the total number of globular clusters may be underestimates.

We have performed extensive tests to study the effect of background galaxies
on our results and the expected amount of contamination. They show that
galaxies and clusters cannot be clearly distinguished from their loci in the 
$r_h - M_V$ and $\mu_{0,V} - r_{c}$ parameter spaces. However, if high resolution 
images are combined with an appropiate colour index like the Washington $(C-T_{1})$,
the level at which the GC sample is contaminated by background galaxies is
about 10\%. Therefore, we expect that our results reflect largely the
physical properties of actual clusters rather than background galaxies.

\begin{acknowledgements}
M.G. thanks S\o ren Larsen for his help with {\it ishape\/} and comments, as 
well as Avon Huxor for the data in Fig.~\ref{fig.MV_rh}. D.G. gratefully acknowledges 
support from the Chilean {\sl Centro de Astrof\'\i sica} FONDAP No. 15010003.
We thank the referee for his/her comments and suggestions which greatly improved
this paper, especially the discussion about the galaxy contamination.
\end{acknowledgements}

\end{document}